\documentclass[12pt]{article}
\usepackage{epsfig,amsfonts}
\begin{document}

\voffset -0.4 true cm
\hoffset 1.3 true cm
\topmargin 0.0in
\evensidemargin 0.0in
\oddsidemargin 0.0in
\textheight 8.7in
\textwidth 6.9in
\parskip 10 pt

\newcommand{\be}{\begin{equation}}
\newcommand{\ee}{\end{equation}}
\newcommand{\bea}{\begin{eqnarray}}
\newcommand{\eea}{\end{eqnarray}}
\newcommand{\beas}{\begin{eqnarray*}}
\newcommand{\eeas}{\end{eqnarray*}}

\font\cmss=cmss10
\def\half{{1 \over 2}}
\def\identity{{\rlap{\cmss 1} \hskip 1.6pt \hbox{\cmss 1}}}
\def\laplace{{\kern1pt\vbox{\hrule height 1.2pt\hbox{\vrule width 1.2pt\hskip
  3pt\vbox{\vskip 6pt}\hskip 3pt\vrule width 0.6pt}\hrule height 0.6pt}
  \kern1pt}}
\def\scriptlap{{\kern1pt\vbox{\hrule height 0.8pt\hbox{\vrule width 0.8pt
  \hskip2pt\vbox{\vskip 4pt}\hskip 2pt\vrule width 0.4pt}\hrule height 0.4pt}
  \kern1pt}}
\def\slash#1{{\rlap{$#1$} \thinspace /}}
\def\roughly#1{\raise.3ex\hbox{$#1$\kern-.75em\lower1ex\hbox{$\sim$}}}
\def\complex{{\hbox{\cmss C} \llap{\vrule height 7.0pt
  width 0.4pt depth -.4pt \hskip 0.5 pt \phantom .}}}
\def\real{{\hbox{\cmss R} \llap{\vrule height 6.9pt width 0.4pt
  depth -.1pt \hskip 0.6 pt \phantom .}}}
\def\integer{{\rlap{\cmss Z} \hskip 1.8pt \hbox{\cmss Z}}}

\font\cmsss=cmss8
\def\C{{\hbox{\cmsss C}}}
\def\bigC{{\hbox{\cmss C}}}
\def\gs{{g^2_{YM}}}
\def\gssb{{g^2_{YM} \sqrt{\beta}}}
\def\gsb{{g^2_{YM} \beta}}
\def\gfb{{g^4_{YM} \beta}}

\begin{titlepage}
\begin{flushright}
{\small BROWN-HET-1233} \\
{\small CU-TP-981} \\
{\small IAS-SNS-99/114} \\
{\small PUPT-1942} \\
{\small hep-th/0007051}
\end{flushright}

\begin{center}

\vspace{1mm}

{\Large \bf Black Hole Thermodynamics from Calculations in
Strongly-Coupled Gauge Theory}

\vspace{2mm}

Daniel Kabat${}^{1,2}$, Gilad Lifschytz${}^{3,4}$\footnotetext[4]{Address
after Aug.~1: Department of Mathematics and Physics, University of
Haifa at Oranim, Tivon 36006, Israel.} and David A. Lowe${}^5$

\vspace{1mm}

${}^1${\small \sl Department of Physics} \\
{\small \sl Columbia University, New York, NY 10027} \\
{\small \tt kabat@physics.columbia.edu}

${}^2${\small \sl School of Natural Sciences} \\
{\small \sl Institute for Advanced Study, Princeton, NJ 08540}

${}^3${\small \sl Department of Physics} \\
{\small \sl Princeton University, Princeton, NJ 08544} \\
{\small \tt gilad@viper.princeton.edu}

${}^5${\small \sl Department of Physics} \\
{\small \sl Brown University, Providence, RI 02912} \\
{\small \tt lowe@het.brown.edu}

\end{center}

\vskip 0.3 cm

\noindent
We develop an approximation scheme for the quantum mechanics of $N$
D0-branes at finite temperature in the 't Hooft large-$N$ limit.  The
entropy of the quantum mechanics calculated using this approximation
agrees well with the Bekenstein-Hawking entropy of a ten-dimensional
non-extremal black hole with 0-brane charge.  This result is in accord
with the duality conjectured by Itzhaki, Maldacena, Sonnenschein and
Yankielowicz.  Our approximation scheme provides a model for the
density matrix which describes a black hole in the strongly-coupled
quantum mechanics.

\end{titlepage}

\section{Introduction}

Properties of black holes in quantum theories of gravity have
intrigued physicists for many years. In recent years much progress has
been made in understanding some of these properties using string
theory. However because string theory is usually formulated
perturbatively about a given spacetime background, it has been
difficult to obtained a unified and complete description of black hole
physics.  In the past few years nonperturbative formulations of string
theory have been proposed in terms of large $N$ gauge theories. These
formulations give new hope for understanding some of these fundamental
issues.

In particular, Maldacena's conjecture gives a promising arena to
answer some of these questions.  Maldacena's conjecture \cite{malda}
relates string theories in anti-de Sitter backgrounds to conformal
field theories in a large-$N$ limit. However there is no free
lunch. In this framework, seemingly obvious questions on the string
theory side are hard to formulate on the gauge theory side.  Moreover,
in the range of parameters where semi-classical string theory may be
used to construct black hole geometries, the dual gauge theory is
strongly coupled. This makes understanding black hole states very
difficult.

One way of dealing with these difficulties is to perform Monte Carlo
simulations of the gauge theory at strong coupling \cite{JanikWosiek,
Ambjorn}.  In the present work we pursue a different approach, in
which we begin with an ansatz for the density matrix which describes
the black hole in the continuum gauge theory.

The version of Maldacena duality to be considered here relates black
holes in ten dimensions with 0-brane charge to supersymmetric gauged
$SU(N)$ quantum mechanics with sixteen supercharges \cite{imsy}. The
metric of the non-extremal black hole is
\bea
\nonumber
ds^2 & = & \alpha'\left[-h(U)dt^2 + h^{-1}(U)dU^2 + {c^{1/2}
(g_{YM}^2 N)^{1/2} \over U^{3/2}} d\Omega_{8}^{2}\right] \\
\label{metric}
h(U) & = & \frac{U^{7/2}}{c^{1/2} (g_{YM}^2 N)^{1/2}}\left(1-\frac{U_{0}^{7}}{U^{7}}\right)
\eea
where $c = 2^7 \pi^{9/2} \Gamma(7/2)$ and $g_{YM}$ is the Yang-Mills
coupling constant.  The horizon of the black hole is at $U=U_0$, which
corresponds to a Hawking temperature
\be
T = (g^2_{YM} N)^{-1/2} \left(U_0 \over 11.1 \right)^{5/2}\,.
\ee
The dual quantum mechanics is to be taken at the same finite
temperature.  The black hole has a free energy, which arises from its
Bekenstein-Hawking entropy \cite{KleTse}.
\begin{equation}
\label{beken}
\beta F = - 2.52 ~ N^2 \left({T \over (g_{YM}^2 N)^{1/3}} \right)^{1.80}
\end{equation}
Duality predicts that the quantum mechanics should have the same free
energy.  The supergravity description is expected to be valid when
the curvature and the dilaton are small near the black hole
horizon. This regime corresponds to the 't Hooft large $N$ limit of
the quantum mechanics, when the dimensionless effective coupling
$g_{YM}^2 N/T^3$ is large.

In this report we describe a set of approximations that can be applied
to the quantum mechanics in the regime of interest.  Using these
techniques we calculate the finite temperature partition function of
the quantum mechanics.  Over a certain range of temperature our
results can be well fit by a power law,
\be
\label{intro:fit}
\beta F \approx {\rm const.} - 2.0 ~ N^2 \left({T \over (g^2_{YM} N)^{1/3}}
\right)^{1.7}\,.
\ee
This is in quite good agreement with the black hole prediction
(\ref{beken}).  We believe this is the first nontrivial direct test of
a strong/weak coupling duality that does not rely on supersymmetric
non-renormalization theorems or special properties of BPS states.
Although in the present paper we are primarily interested in the
thermodynamics of the quantum mechanics, our approximation scheme
should also be useful for addressing questions about the spacetime
structure of the black hole, perhaps along the lines of \cite{KL1,KL2}.

The basic idea is to treat the ${\cal O}( N^2)$ degrees of freedom of
the quantum mechanics as statistically independent, using a type of
mean field approximation.  This assumption is motivated by the overall
$N^2$ dependence of the free energy (\ref{beken}).  The approximation
involves constructing a trial action $S_0$ from the full action $S$.
All quantities can then be systematically computed as an expansion in
powers of $S - S_0$.  The parameters in the trial action are fixed by
solving a truncated version of the Schwinger-Dyson equations of the
quantum mechanics.  This procedure can be viewed as re-summing an
infinite number of Feynman diagrams.  Since we are interested in
large-$N$ behavior, we will only resum planar diagrams.  Thus, in our
approximation, the overall $N^2$ factor in the free energy
(\ref{intro:fit}) and the appearance of $g_{YM}^2$ only in the
combination $g_{YM}^2 N$ is guaranteed.  The crucial test of the
approximation is to obtain the correct dependence of the
thermodynamics on the effective dimensionless coupling $g_{\rm eff}^2
= g_{YM}^2 N/T^3$.

The sort of approximation that we are considering has several
attractive features, which we regard as {\it a priori} reasons to
believe that it captures some of the essential physics of the
strongly-coupled quantum mechanics.
\begin{itemize}
\item As mentioned above, the approximation automatically respects 't
Hooft large-$N$ counting.
\item The approximation partially respects the symmetries of the
problem.  More precisely, it respects all symmetries which act linearly on
the fields.  Thus our trial action will have ${\cal N} = 2$ supersymmetry and
$SO(2) \times SO(7)$ rotational symmetry (out of the underlying ${\cal
N} = 16$ supersymmetry and $SO(9)$ rotational symmetry).
\item The approximation is non-perturbative in the Yang-Mills coupling
constant, and self-consistently cures the infrared divergences
present in conventional perturbation theory.
\end{itemize}

Before continuing there is one other issue we would like to comment
on.  The partition function of the full quantum mechanics contains an
infrared divergence from the regions in moduli space when the
D0-branes are far apart. This leads to a divergent contribution to the
entropy with an overall coefficient ${\cal O}(N)$.  From the
supergravity point of view, this corresponds to a thermal gas of
gravitons. This divergence may be regulated by putting the system in a
finite box. The black hole entropy which is ${\cal O}(N^2)$ can easily
be made to dominate over the ${\cal O}(N)$ contribution. Our mean
field approximation computes the ${\cal O}(N^2)$ piece by design, and
this infrared divergence does not make an appearance.

\section{Gaussian Approximation for 0-brane Quantum Mechanics}

In this section we sketch the application of the Gaussian
approximation \cite{kl} to gauged $SU(N)$ supersymmetric quantum
mechanic with sixteen supercharges. Further details will appear in
\cite{kll}.

One key requirement is that supersymmetry, although softly broken by
the finite temperature of the black hole, should not be broken
explicitly by the approximation.  To avoid such explicit breaking we
adopt an unconstrained superfield formulation, in which supersymmetry
acts linearly on the fields.  This ensures that we recover exact
supersymmetry in the zero temperature limit.  We will use an ${\cal N}
= 2$ superspace so that only an $SO(2) \times G_2$ subgroup of the
$SO(9)$ R-symmetry is manifest.  First let us recall the superspace
and supermultiplets that we are going to use; for more details on
notation see \cite{kl}.

With ${\cal N} = 2$ supersymmetry we have an $SO(2)$ R-symmetry, with
spinor indices $\alpha,\beta = 1,2$ and vector indices $i,j = 1,2$.
The $SO(2)_R$ Dirac matrices $\gamma^i_{\alpha\beta}$ are real,
symmetric, and traceless.  Given two spinors $\psi_\alpha$ and
$\chi_\alpha$, besides the invariant $\psi_\alpha \chi_\alpha$, one
can construct a second invariant which we denote
\[
\psi^\alpha \chi_\alpha = {i \over 2} \epsilon_{\alpha\beta} \psi_\alpha \chi_\beta\,.
\]
${\cal N} = 2$ superspace has coordinates $(t,\theta_\alpha)$ where
$\theta_\alpha$ is real.  We denote $\theta^2 = \theta^\alpha
\theta_\alpha$, and define the supercovariant derivative
\begin{equation}
D_\alpha  =  {\partial \over \partial \theta_\alpha} - i \theta_\alpha {\partial \over
\partial t}~.
\end{equation}
The simplest representation of supersymmetry is a real scalar superfield
\be
\Phi = \phi + i \psi_\alpha \theta_\alpha + f \theta^2
\ee
containing a scalar field $\phi$, its superpartner $\psi_\alpha$, and
an auxiliary field $f$.  The gauge superfield denoted by
$\Gamma_{\alpha}$ has an expansion in `linear' components as
\be
\Gamma_\alpha = \chi_\alpha + A_0 \theta_\alpha + X^i \gamma^i_{\alpha\beta} \theta_\beta
+ d \epsilon_{\alpha\beta} \theta_\beta + 2 \epsilon_{\alpha\beta} \lambda_\beta \theta^2 \,.
\ee
The fields $X^i$ are physical scalars, while $\lambda_\alpha$ are
their superpartners, $d$ is an auxiliary boson, $\chi_\alpha$ are
auxiliary fermions, and $A_0$ is the 0+1 dimensional gauge field.

To construct our action we introduce a collection of seven adjoint
scalar multiplets $\Phi_a$ transforming in the ${\bf 7}$ of
$G_{2}\subset SO(9)$.  The ${\cal N}=16$ supersymmetric $SU(N)$ gauged
quantum mechanics action is
\[
S_{\rm sym}  =  {1 \over g^2_{\rm YM}} \int dt d^2\theta \, {\rm Tr} \left( - {1 \over 4}
\nabla^\alpha {\cal F}_i \nabla_\alpha {\cal F}_i - \frac{1}{2} \nabla^{\alpha}
\Phi_{a} \nabla_{\alpha}\Phi_{a} - \frac{i}{3} f_{abc}\Phi_{a}[\Phi_{b},\Phi_{c}]\right)
\]
where ${\cal F}_{i}$ is the field strength constructed from the gauge
multiplet, $\nabla_{\alpha} = D_{\alpha} +\Gamma_{\alpha}$, and
$f_{abc}$ is a suitably normalized totally antisymmetric
$G_{2}$-invariant tensor.

We impose the supersymmetric gauge condition $D^{\alpha}
\Gamma_{\alpha}=0$, which sets $\partial A_0/\partial t = 0$, $d=0$
and $\lambda = \frac{1}{2} \partial \chi / \partial t$.  This is a
convenient gauge fixing, as this gauge condition helps make the
approximation compatible with Ward identities \cite{kll}.  To the SYM
action we must add the corresponding ghost action (but no gauge fixing
term)
\[
S_{\rm ghost} = {1 \over g^2_{\rm YM}} \int dt d^2\theta \, {\rm Tr} \left(
 D^\alpha \bar{C} \nabla_\alpha C \right)
\]
where the ghost superfield $C$ is a complex scalar superfield with
Grassmann statistics.

We are interested in the finite temperature behavior of the quantum
mechanics.  As usual we compactify the Euclidean time coordinate on a
circle of circumference $\beta$, which is identified with the inverse
temperature.  The bosonic fields have integer mode expansions, while
the fermions have half integer modes; for example we write
\[
X^i(\tau) = {1 \over \sqrt{\beta}} \sum_{l \in {\mathbb Z}} X_l^i e^{i 2 \pi l \tau / \beta}\,.
\]
Note that in Euclidean space the zero mode of the gauge field, which
we denote $A_{00}$, survives as a physical degree of freedom
(fluctuations in $A_0$ are eliminated by our gauge condition).

Our strategy is to construct a trial action $S_0$, which we use as an
approximation to the full action $S= S_{\rm sym} + S_{\rm ghost}$.
The free energy can then be calculated as an expansion in powers of
$S-S_0$; note that such an expansion is non-perturbative in the
original Yang-Mills coupling $g^2_{YM}$.  Throughout this paper we use
the following expression for the free energy in this
scheme.\footnote{This quantity can be identified with the two-loop 2PI
effective action of \cite{CJT}.}
\begin{equation}
\label{variat}
\beta F \approx \beta F_{0} + \langle S-S_{0} \rangle_{0} - 
\frac{1}{2}\langle S_{III}^{2}\rangle_{\C,0}
\end{equation}
Here $\beta F_0$ is the free energy of the trial action, and $\langle
\cdot \rangle_0$ denotes an expectation value computed using $S_0$.
Also $S_{III}$ refers to cubic terms in the original SYM plus ghost
action, and the subscript $\bigC$ denotes a connected correlation
function.  It is straightforward, though tedious, to compute higher
order terms in the expansion of $\beta F$.  In principle this could be
used as a check on the validity of the approximation.

We make the following ansatz for the trial action:
\be
\label{trial}
S_{0}=-\frac{N}{\lambda}{\rm Tr}(U +U^{\dagger})+
\sum_{l,i} \frac{1}{2 \sigma_{l}^{2}}{\rm Tr}(X^{i}_{l} X^{i}_{-l})  + 
\sum_{l,a} \frac{1}{2 \Delta_{l}^{2}}{\rm Tr}(\phi^{a}_{l}
\phi^{a}_{-l}) + \cdots\,.
\ee
Here all fields (except the gauge field) appear in Gaussian form.
Indeed this is the most general Gaussian action which is quadratic in
the fundamental fields.  This means that the trial action can respect
all symmetries which act linearly on the fundamental fields.

The gauge field must be treated in a special way, owing to its
periodicity properties.  To do this we have introduced the timelike
Wilson loop operator $U$, which can be expressed in terms of the gauge
zero mode.
\[
U = P e^{i \int_0^\beta A_0} = e^{i\sqrt{\beta}A_{00}}
\]
This makes it manifest that at finite temperature $A_0 \sim A_0 + 2
\pi/\beta$ is periodic.  As a trial action for the gauge field we have
adopted the unitary one plaquette model action.  As $\lambda$ varies
the trial action goes through a Gross-Witten phase transition at
$\lambda = 2$ \cite{GrossWitten}.

The key step in the approximation is to find a closed set of ``gap''
equations for the dressed propagators $\sigma_l^2,\, \Delta_l^2,\,
\ldots$ appearing in (\ref{trial}).  Again, the gauge field must be
treated as a special case.  All other propagators are obtained by
demanding stationarity of the estimate (\ref{variat}) for the free
energy.  Up to contributions from the gauge field, it can be shown
this procedure correctly re-sums all one-loop self-energy corrections
to the propagators.

The gap equation for $\lambda$ is obtained from the Schwinger-Dyson
equation for $\langle U\rangle$ that arises from the change of
variables $U\to g U$ with $g \in SU(N)$. Demanding that this equation
hold with respect to the one-plaquette measure yields
\be
\label{gaugeSD}
\langle {\rm Tr} U \rangle_0 = \frac{1}{\beta^{3/2}} {\rm Tr}
\left\langle U \left( \frac{\delta S}{\delta A_{00}} - \frac{1}{2} \frac{\delta
  S_{III}^2}{\delta A_{00}}  \right) \right\rangle_{C,0} ~.
\ee
This equation resums one-loop corrections to the Wilson loop, in the
same sense that stationarizing (\ref{variat}) resums one-loop corrections to the
propagators.  At large $N$ the terms on the right hand side factorize
into a gauge field correlator times matter field correlators; the
terms involving the gauge fields may be computed using the results of
\cite{GrossWitten}.

As (\ref{variat}), (\ref{gaugeSD}) are somewhat lengthy expressions we
will not present them here.  Rather we report on the solution to these
equations in the next section.

\section{Numerical Results}

The gap equations can be solved numerically, using the methods
discussed in Appendix B of \cite{kl}.  The basic strategy is to start
at high temperature, where the gap equations can be solved
semi-analytically, then use Newton-Raphson to solve the gap equations
at a sequence of successively lower temperatures. As mentioned in the
introduction, $g_{YM}^2$ and $N$ only appear in the combination
$g_{YM}^2 N/T^3$. Henceforth we choose units for $T$ which effectively
scales $g_{YM}^2 N$ to $1$.

In principle the resulting Gaussian action contains a great deal of
information about correlation functions in the quantum mechanics.  But
in this section we will just concentrate on the behavior of three
basic quantities: the free energy, the Wilson loop, and the mean size
of the state.

At high temperature, where the gauge theory is weakly coupled, we find
that the free energy of the system is
\be
\beta F = 6 \log \beta + {\cal O}(1)\,.
\ee
This result can be obtained analytically: the gap equations are
dominated by the bosonic zero modes, and the free energy is dominated
by $\beta F_0$.

In general, for a weakly-coupled theory in $0+1$ dimensions, one would
expect the free energy to behave like $\log \beta$.  But note that,
even though the gauge theory is weakly coupled at high temperature,
the perturbation series is afflicted with IR divergences.  Thus, to
determine the coefficient of the logarithm (which depends on the value
of the dynamically generated IR cutoff) one must re-sum part of the
perturbation series.  This is a well-known phenomenon in finite
temperature field theory \cite{DolanJackiw}.  In any case, we expect
{\em a priori} that the Gaussian approximation gives good results in
the high temperature regime.

\begin{figure}
\epsfig{file=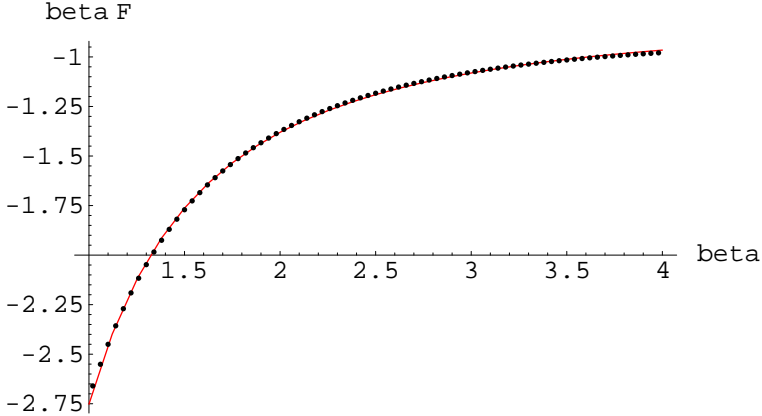}
\caption{The solid curve is the power law fit (\ref{results:fit})
for $\beta F$.  The data points are calculated from numerical
solutions to the gap equations.}
\end{figure}

As the temperature is lowered the behavior of the free energy changes:
at $\beta \approx 0.7$ we find that it begins to roll over and fall
off as a non-trivial power of the temperature.  In the range $1 <
\beta < 4$ the numerical results for the free energy are well fit by
\be
\label{results:fit}
\beta F \approx - 0.79 - 2.0 ~ \beta^{-1.7}\,.
\ee
This fit to the numerical results is illustrated in Fig.~1.  Note that
supersymmetry is crucial in making such power-law behavior possible.
Without supersymmetry the free energy would behave as $\beta F \approx
\beta E_0$ in the low temperature regime $\beta > 1$, where $E_0$ is
the ground state energy of the system.

We obtained (\ref{results:fit}) by performing a Levenberg-Marquardt
nonlinear least-squares fit to 75 numerical calculations of the free
energy, carried out in the temperature range $1 \leq \beta \leq 4$. To
estimate the uncertainty in the best fit parameters we varied the
window of $\beta$ over which the fit was performed (fitting over the
ranges $2<\beta<4$ and $1<\beta<3$), which leads to: $-0.79 \pm 0.06$,
$-2.0 \pm 0.1$ and $-1.7\pm 0.2$.

It is quite remarkable that the power law (\ref{results:fit}) is in
excellent agreement with the semiclassical black hole prediction
\cite{imsy}
\begin{equation}
\beta F = - 2.52 ~ \beta^{-1.80}\,.
\end{equation}
The exponents differ by 6\% while the coefficients of the
power-law differ by 26\%.  (An additive constant appears in the
Gaussian approximation for the free energy.  We will generally ignore
this `ground state degeneracy', since it seems to be an artifact of
the Gaussian approximation when applied to systems with a continuous
spectrum.  Similar behavior was noted in \cite{kl}.)

As we go to still lower temperatures, we find that the energy
$\partial (\beta F)/\partial\beta$ calculated in the Gaussian
approximation begins to drop below the energy of the black hole.  In
fact the Gaussian energy becomes negative around $\beta = 5.8$.
Ultimately, as $\beta \rightarrow \infty$, the Gaussian energy does
asymptote to zero, as required by the ${\cal N} = 2$ supersymmetry
which is manifest in the approximation.  But a negative energy clearly
reflects some problem with the approximation.

Fortunately, we can be rather precise about exactly where the
approximation is going wrong: the difficulty is with the
Schwinger-Dyson gap equation we have been using to fix the value of
the one-plaquette coupling $\lambda$.  Although we do not know how to
write down a better gap equation for $\lambda$, we can give a {\em
prescription} for fixing $\lambda$, that will allow us to obtain
reasonable results at much lower values of the temperature. This may
be regarded either as a check on our understanding of why the
approximation is breaking down, or as a way of building a model for
the black hole that can be used at lower temperatures.  Our
prescription for fixing $\lambda$ is simply that, when $\beta > 2.5$
(the midpoint of our range $1 \le \beta \le 4$), we choose $\lambda$
so that the free energy is given by (\ref{results:fit}).

\begin{figure}
\epsfig{file=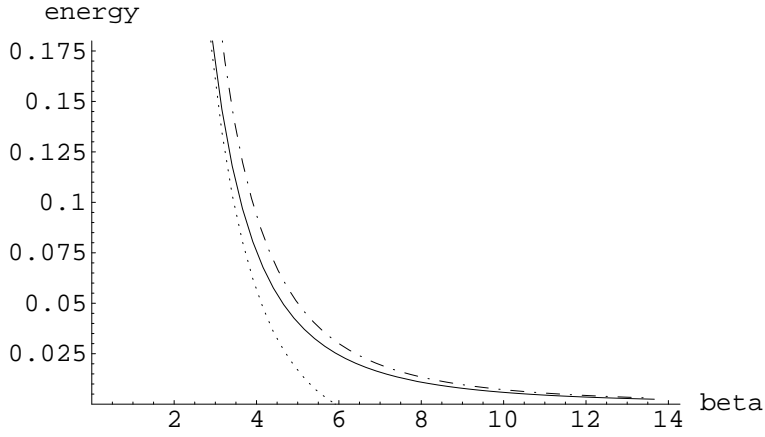}
\caption{Energy vs.~$\beta$.  For $\beta >2.5$ fixing $\lambda$ by
fitting $\beta F$ to a power law leads to the solid middle line, while
the Schwinger-Dyson gap equation for lambda leads to the lower dashed 
line.  The upper dot-dashed line is the semiclassical energy of the
black hole.}
\end{figure}

\begin{figure}
\epsfig{file=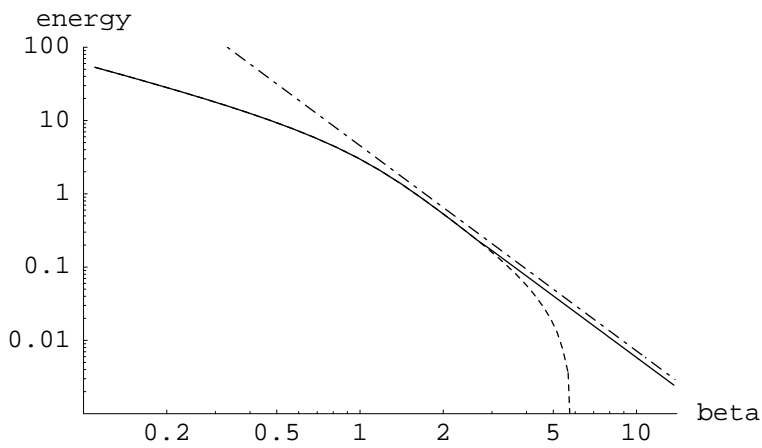}
\caption{Energy vs.~$\beta$.  Same as Fig.~2, but plotted on a
$\log$--$\log$ scale.}
\end{figure}

\begin{figure}
\epsfig{file=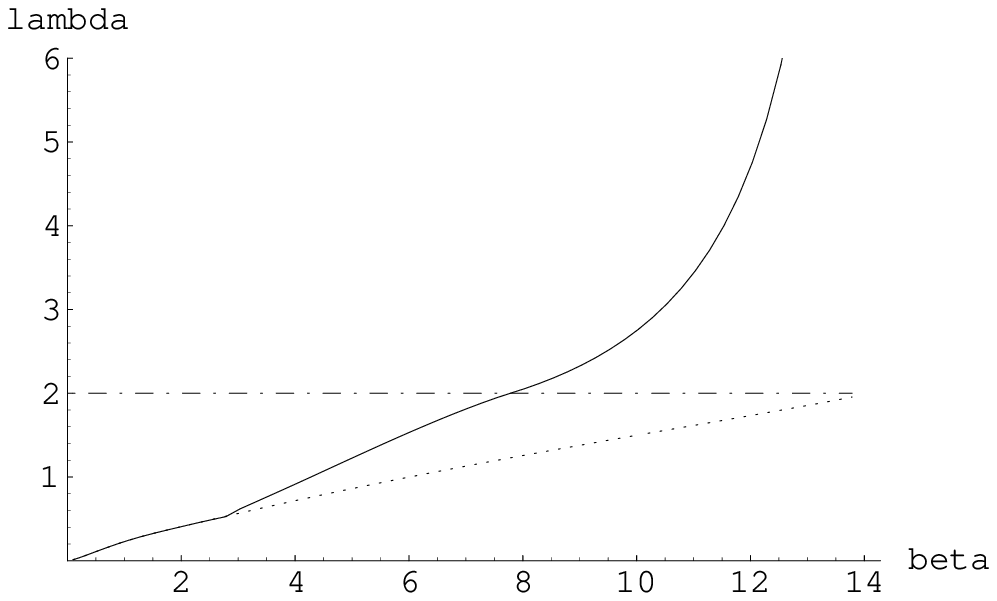}
\caption{The one-plaquette coupling $\lambda$ vs.~$\beta$.  The
Gross-Witten transition occurs when $\lambda = 2$.  For $\beta < 2.5$ we
use the Schwinger-Dyson gap equation to determine $\lambda$.  For
$\beta > 2.5$ the Schwinger-Dyson gap equation gives the dashed 
line, while fitting $\beta F$ to a power law gives the solid 
line.}
\end{figure}

The energy $E = \partial (\beta F) / \partial \beta$ calculated with
this prescription is shown in Figs.~2 and 3.  The behavior of
$\lambda$ given by this prescription is shown in Fig.~4.  Note that
$\lambda$ increases monotonically with $\beta$.  A Gross-Witten phase
transition takes place when $\lambda = 2$; this value is reached at
$\beta = 7.8$.  Thus a phase transition takes place as the system
moves into the supergravity regime \cite{kl}.

By adopting the prescription of fitting $\beta F$ to a power law, we
cannot say anything about the order of the phase transition. 
If one takes the Schwinger-Dyson
result for $\lambda$ seriously, then the Gross-Witten transition
occurs at $\beta = 14.2$, and is second order (the second derivative
of the free energy drops by $0.01$ in crossing the
transition).

Our prescription for choosing $\lambda$ begins to break down around
$\beta = 14$, as we find that $\lambda$ rapidly diverges as $\beta$
approaches 14.  By itself, this is not necessarily a problem:
infinite $\lambda$ simply means that the Wilson loop is uniformly
distributed over $U(N)$.  But unfortunately, we do not have a good
prescription for continuing past this temperature.  Evidently some of
the other gap equations (not just the gap equation for $\lambda$)
start to break down at this point.  Note that this breakdown does not
occur until well into the strong coupling regime, as an inverse
temperature $\beta = 14$ corresponds to an effective gauge coupling
$g_{\rm eff}^2 = \beta^3 \approx 3 \times 10^3$.

Finally, let us comment on the average `size' of the state.  In our
approximation the scalar fields $X^i(\tau)$ and $\phi^a(\tau)$ are
Gaussian random matrices, and their eigenvalues obey a Wigner
semi-circle distribution.  We can define the size of the state in terms
of the quantities
\bea
\label{radius}
R^2_{\rm scalar}  &=&  {1 \over N} {\rm Tr} \langle
\left(X^i(\tau)\right)^2 \rangle_0 ~, \nonumber \\
R^2_{\rm gauge} & =&  {1 \over N} {\rm Tr} \langle
\left(\phi^a(\tau)\right)^2 \rangle_0 \,.
\eea
The radius of the Wigner semi-circle, given by $2 \sqrt{R^2}$, is shown
in Fig.~5.  Note that the radius stays fairly constant in the region
corresponding to the black hole.  However, because the superfield
formalism we are using does not respect the full $SO(9)$ invariance,
the radius measured in the scalar multiplet directions is not the same
as the radius measured in the gauge multiplet directions.  At $\beta = 14$
we find
\[
2 R_{\rm scalar} = 1.81 \qquad\quad 2 R_{\rm gauge} = 0.80\,.
\]
This shows that, as expected, the trial action does not respect the
underlying $SO(9)$ invariance.  Nonetheless, the trial action may
provide a useful approximate description of the black hole density
matrix in the supergravity regime.

\begin{figure}
\epsfig{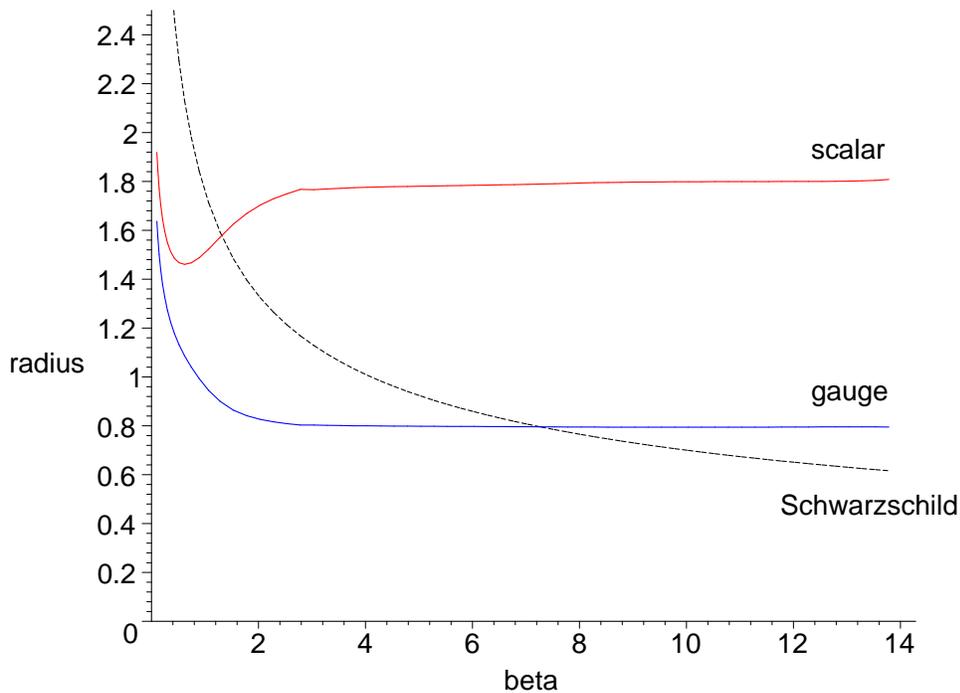}
\caption{Range of eigenvalues (radius of the Wigner semi-circle)
vs.~$\beta$.  The upper solid red curve is for the scalar fields in the
scalar multiplets; the lower solid blue curve is for the scalar fields in
the gauge multiplet.  The dashed black curve is the Schwarzschild
radius of the black hole.  These results were calculated with $\beta
F$ fit to a power law for $\beta >2.5$.}
\end{figure}

In Fig.~5 we have also plotted the Schwarzschild radius of the black
hole\footnote{Our Higgs fields are related to radial position by $X =
r / 2 \pi \alpha'$, while ref.~\cite{imsy} sets $U = r / \alpha'$.}
$U_0/2\pi=1.76~\beta^{-2/5}$.  Note that, as the
temperature decreases, the Schwarzschild radius becomes much smaller
than the radius of the eigenvalue distributions.  It seems appropriate
to identify the radius of the eigenvalue distributions with the size
of the region $U \ll (g^2_{\rm YM} N)^{1/3}$ in which 10-dimensional
supergravity is valid \cite{imsy}.  The Gaussian approximation should
provide a good laboratory for studying the way in which the horizon of
the black hole can be detected in the dual quantum mechanics, perhaps
along the lines suggested in \cite{KL1,KL2}.

\section{Discussion}

To summarize, we have presented an ansatz for a trial action which
captures some of the behavior of 0-brane quantum mechanics at large
$N$ and strong coupling.  The parameters appearing in the trial action
are chosen according to a set of gap equations which resum an infinite
set of planar diagrams.  The approximation automatically respects 't
Hooft large-$N$ counting, and also partially respects the
supersymmetries and R-symmetries of the quantum mechanics.

Our main result is that we obtained a non-trivial power law for the
free energy, which is in remarkably good agreement with the black hole
prediction.  The power law was obtained by fitting numerical results
calculated in the range $1 < \beta < 4$; we would like to emphasize
that in this range of temperature the approximation involves no
arbitrary or adjustable parameters.  We also gave a prescription for
fixing the expectation value of the Wilson loop, which allowed us to
extend our results to $\beta = 14$.  This is well into the strong
coupling regime, where the supergravity description of the system
should be valid.

Although our prescription for fixing the Wilson loop cannot be
regarded as entirely satisfactory, it does let us build a model for
the black hole density matrix in the strongly-coupled quantum
mechanics.  This model could be a basis for further studies of black
hole properties.  It would be particularly interesting to address
questions of spacetime locality and causality from the gauge theory
point of view, perhaps along the lines suggested in \cite{KL1,KL2}.

\bigskip
{\bf Acknowledgments}

DK wishes to thank Rutgers University and New York University for
hospitality, and Emil Martinec and Mark Stern for valuable
discussions.  The work of DK is supported by the DOE under contract
DE-FG02-90ER40542 and by the generosity of Martin and Helen Chooljian.
GL would like to thank the Aspen Center for Physics for hospitality,
and Vipul Periwal for useful discussions.  The work of GL is supported
by the NSF under grant PHY-98-02484.  D.L. wishes to thank the Aspen
Center for Physics, the Abdus Salam International Centre for
Theoretical Physics and the Erwin Schroedinger Institute program in
Duality, String Theory and M-Theory for hospitality during the course
of this research.  The research of D.L. is supported in part by DOE
grant DE-FE0291ER40688-Task A.


\end{document}